# A fast algorithm to find reduced hyperplane unit cells and solve N-dimensional Bézout's identities

Authors

**Cyril Cayron**[a]*

[a]Laboratory of Thermo Mechanical Metallurgy (LMTM), PX Group Chair, EPFL, Rue de la Maladière 71b, Neuchâtel, 2000, Switzerland

Correspondence email: cyril.cayron@epfl.ch

**Synopsis**   The paper explains the method to determine a short unit cell attached to any hyperplane given by its integer vector **p**. Equivalently, it gives all the solutions of the *N*-dimensional Bézout's identity associated with the coordinates of **p**.

**Abstract**   An algorithm is proposed to find a short solution and the affine set of solutions of Bézout's identities. The first part of the paper is devoted to the column-constrained unimodular matrix problem. which consists in finding an integer matrix such that the absolute value of its determinant is 1 and the values of its last column are fixed. In a second part, a recursive algorithm that determines some solutions to the *N*-dimensional Bézout's identity is presented. The third part shows how to combine the two previous algorithm and uses parallel projections to determine a unit cell attached to any integer hyperplane **p**. This unit cell can be then shorten by lattice reduction with LLL or cubification method. The vectors of the unit cell form the affine space of all the solutions of the *N*-dimensional Bézout's identity based on the vector **p**.

## 1. Introduction

This study was initiated by the study made by Gorfman (2020) on what we will call the "column-constrained unimodular matrix problem" which consists in determining an integer matrix **M** such that its determinant is $\pm 1$ and the last column is equal to a fixed vector **t**. Since Gorfman proposed a solution that seems reliable but slow, we have looked for another and more effective algorithm. The algorithm we propose is based on 2D Bézout's identity; it is detailed in §2. In parallel, we developed in §3 an algorithm to determine some solutions to the *N*-dimensional Bézout's identity. In §4, we explain how to combine the algorithms of §2 and §3 to attach a unit cell to any hyperplane given by its normal (reciprocal) vector, which could be useful in crystallography and geometry. Equivalently, the algorithm gives the infinite set of solutions of *N*-dimensional Bézout's identity, which could have application in number theory and algebra. It is also shown how the unit cell can be further reduced thanks to lattice reduction algorithms (LLL or cubification); i.e. how to determine an equivalent unit cell with shorter and more orthogonal vectors.

In this paper, $u_i$ designs the $i^{th}$ coordinate of a vector **u**. Sometimes, the notation $\mathbf{u}_{(i)}$ will be also equivalently used. It should not be confused with $\mathbf{u}_i$ that is the $i^{th}$ vector in a set of vectors $\{\mathbf{u}_i\}$. The coordinates of a vector **u** are written in column and those of the vector noted $\mathbf{u^t}$ are in line. From a crystallographic point of view, column and line vectors belong to direct and reciprocal spaces, respectively. The matrix multiplication notation is adopted. It means that even a "simple" scalar product $\mathbf{p} \cdot \mathbf{u} = \sum_i p_i u_i$ is written $\mathbf{p^t u}$ where "$\mathbf{p^t}$" means "transpose of **p**". Please note that a list of vectors are represented by a matrix formed by the vector coordinates are written in columns, whereas in the companion paper about lattice reduction (Cayron, 2021b), the vectors are written in row in order to respect the usual custom in lattice reduction problems.

## 2. Algorithm to solve the column-constrained unimodular matrix problem

### 2.1. Case where one of the coordinates of t is $\pm 1$

In the column-constrained unimodular matrix problem, there is a simple and immediate solution if the $N^{th}$ coordinate of **t** is $\pm 1$. In that case, any triangular matrix **M** with 1 in the diagonal and with **t** as last column checks the condition $det(\mathbf{M}) = 1$. If one of the coordinates of **t** is 1 but in a position $i < N$, then a simple matrix of permutation **P** is sufficient to recalculate the matrix **M**. The example used by Gorfman (2020) with the vector **t** of coordinates $[-1,4,2]$ enters in this category. A direct solution is

$$\mathbf{M} = \begin{bmatrix} 0 & 0 & -1 \\ 0 & 1 & 4 \\ 1 & 0 & 2 \end{bmatrix}$$

We will not give more details here because the solutions are actually included in the more general method based on Bézout's identity explained as follows.

### 2.2. Case where t has coprime coordinates

With $N = 2$ the general solution is indeed given by the classical 2D Bézout's identity. If we note $\mathbf{t} = \begin{bmatrix} a \\ b \end{bmatrix}$, there is a solution if and only if the integers $a$ and $b$ are coprime, and the solution is simply $\mathbf{M} = \begin{bmatrix} u & a \\ v & b \end{bmatrix}$, where $u$, $v$ are the Bézout numbers associated with $a$, $b$, i.e. solutions of the equation $au + bv = 1$. If $a$ and $b$ are not coprime, the determinant of any matrix **M** with **t** in last column would be a multiple of gcd($a,b$), the greatest common divisor of $a$ and $b$, and thus cannot be equal to $\pm 1$. A fast and well-known algorithm to determine 2D Bézout numbers is based on iterative Euclidean division (Wikipedia, 2021a).

Now, we consider the case where $N > 2$ and the vector **t** has its two last coordinates $t_{N-1} = a$ and $t_N = b$ that are coprime numbers. A direct solution is the matrix **M** made of two blocks, the top right one is the $(N-2) \times (N-2)$ identity matrix, and the bottom left one is $\begin{bmatrix} u & a \\ v & b \end{bmatrix}$ where $u$, $v$ are the Bézout numbers associated with $a$, $b$. If the two coprime coordinates of vector **t**, $a$ and $b$ are not the last ones and are in positions $i$ and $j$, respectively, the permutation matrices $\mathbf{P}(i, N-1)$ and $\mathbf{P}(j, N)$ can be used to come back to the previous case. We recall that a permutation matrix $\mathbf{P}(i, j)$ is a $N \times N$ identity matrix, except for the line $i$ for which 1 is written in the column $j$, and for the column $j$ where 1 is written in the line $i$. Permutation matrices are unimodular matrices and are equal to their inverse. The unimodular matrix $\mathbf{P} = \mathbf{P}(i, N-1).\mathbf{P}(j, N)$ is such that the vector $\mathbf{P.t}$ has for last coordinates the coprime numbers $a$ and $b$. We thus came back to the case $t_{N-1} = a$ and $t_N = b$. If we call **M** the two-block solution of this case, the solution of the problem is given by the matrix $\mathbf{P}^{-1}.\mathbf{M}$. Please note that $\mathbf{P}^{-1} = \mathbf{P}(j, N).\mathbf{P}(i, N-1) \neq \mathbf{P}$. The cases treated in this section include the case mentioned in §2.1.

With **t** of coordinates $[1551, -540, 67, -102, 2140, -277, 32, 366, 450, 1532]$ used as example, the algorithm gives immediately a solution:

$$\mathbf{M} = \begin{bmatrix} 0 & 0 & 0 & 0 & 0 & 0 & 0 & 0 & -463 & 1551 \\ 0 & 1 & 0 & 0 & 0 & 0 & 0 & 0 & 0 & -540 \\ 0 & 0 & 0 & 0 & 0 & 0 & 0 & 0 & -20 & 67 \\ 0 & 0 & 0 & 1 & 0 & 0 & 0 & 0 & 0 & -102 \\ 0 & 0 & 0 & 0 & 1 & 0 & 0 & 0 & 0 & 2140 \\ 0 & 0 & 0 & 0 & 0 & 1 & 0 & 0 & 0 & -277 \\ 0 & 0 & 0 & 0 & 0 & 0 & 1 & 0 & 0 & 32 \\ 0 & 0 & 0 & 0 & 0 & 0 & 0 & 1 & 0 & 366 \\ -1 & 0 & 0 & 0 & 0 & 0 & 0 & 0 & 0 & 450 \\ 0 & 0 & 1 & 0 & 0 & 0 & 0 & 0 & 0 & 1532 \end{bmatrix}$$

**2.3. Case where t has no coprime coordinates**

Now, let us consider the rarer cases in which none of the pairs $(t_i, t_j)$ of coordinates of **t** are coprime despite the fact that the set of coordinates of **t** is coprime (as mentioned previously, if it were not, there would not be solution to the problem). One says that the set of integers $\{t_i, i = 1, ..., N\}$ is coprime but not pairwise coprime. A classical example of a coprime but not pairwise coprime set is $\{6, 10, 15\}$. Let us recall that in large dimension $N$ the probability that a set of integers that is coprime but not pairwise coprime is very small because the probability that two random integers are coprime is quite high; it is equal to $\frac{1}{\zeta(2)} = \frac{6}{\pi^2} \approx 61\%$, where $\zeta$ refers to the Riemann zeta function (Wikipedia, 2021b). The exact calculation of the probability for a set of $N$ integers to be coprime but not pairwise coprime as function of $N$ is however not straightforward and clearly beyond the scope of the present study. Even if rare, these cases can be solved as follows. We consider the two first coordinates $t_1 = a$ and $t_2 = b$ of the vector **t** (any pair of coordinates would also work). As $a$ and $b$ are not coprime, they can be written $a = xy$ and $b = yz$, where $x$, $y$, $z$ are three integers and $y = \gcd(a,b) > 1$. It is important to note here that there is at least another coordinate $t_i$ with $i > 2$, that cannot be divided by $y$, because if it were not so, $\{t_i\}$ would not be coprime. We call $(u, v)$ the Bézout numbers associated with $(a, b)$, or equivalently to $(x, z)$. We also call $(\alpha, \beta)$ the Bézout numbers associated with $(u, v)$. The determinant of the matrix $\mathbf{B} = \begin{bmatrix} u & v \\ -\beta & \alpha \end{bmatrix}$ is 1, and $\mathbf{B} . \begin{bmatrix} xy \\ yz \end{bmatrix} = \begin{bmatrix} y \\ ky \end{bmatrix}$, with $k \in \mathbb{Z}$. We build the $N \times N$ matrix $\mathbf{M}$ from the $2 \times 2$ block $\mathbf{B}$ and from the identity matrix of dimension $N - 2$. The first coordinate $(\mathbf{B}.\mathbf{t})_{(1)}$ of the new vector $\mathbf{B}.\mathbf{t}$ is coprime with at least one of the coordinates $(\mathbf{B}.\mathbf{t})_{(i)}$ with $i > 2$. It means that the method described in the previous paragraph can be applied to calculate a matrix $\mathbf{M}$ such that $det(\mathbf{M}) = 1$ and such that its last column is the vector $\mathbf{B}.\mathbf{t}$. The matrix $\mathbf{B}^{-1}.\mathbf{M}$ is then such that its determinant is also 1 and its last column is $\mathbf{t}$. As the determinant of $\mathbf{B}$ is 1, $\mathbf{B}^{-1}$ is the adjugate of $\mathbf{B}$, and is thus an integer matrix. Consequently, $\mathbf{B}^{-1} \mathbf{M}$ is also an integer matrix; it is the solution of the problem. The algorithm is effective and fast, whatever the dimension $N$ of the vector $\mathbf{t}$. We just give an example with the classical set of coprime but not coprime coordinates [6,10,15]. The algorithm gives immediately a solution (the vectors are written in columns):

$$\mathbf{M} = \begin{bmatrix} 1 & 0 & 6 \\ 2 & 1 & 10 \\ 0 & -7 & 15 \end{bmatrix}$$

Let us build another example with a vector $\mathbf{t}$ of coordinates $[-42, 10, 15, -30, 6]$. A solution is

$$\mathbf{M} = \begin{bmatrix} 0 & 4 & 0 & 1 & -42 \\ 0 & -1 & 0 & 0 & 10 \\ 0 & 0 & 0 & -7 & 15 \\ -1 & 0 & 0 & 0 & -30 \\ 0 & 0 & 1 & 0 & 6 \end{bmatrix}$$

It is important to note that the algorithm gives only one solution $\mathbf{M}$, but that there is actually an infinite set of solutions, even in the simple case of $N = 2$ because the Bézout numbers are not unique. Indeed, if $u$, $v$ are the Bézout numbers associated with $a$, $b$, i.e. solutions of the equation $au + bv = 1$, then $u - kb$, $v + ka$ are also solutions. The set of solutions form a row that is perpendicular to the vector $[a, b]$ at a unit distance from the origin. It is an affine space of dimension $N - 1 = 1$. This geometrical representation of Bézout number will be used in the next sections.

### 3. N-dimensional Bézout's identity

In this section, we present two methods to calculate some solutions of $N$-dimensional Bézout's identity. Given a set of integers $\{p_i, i = 1, ..., N\}$ we look for another set of integers $\{u_i, i = 1, ..., N\}$ such that $\sum_{i=1}^{N} p_i u_i = 1$. In other words, given an integer vector $\mathbf{p}$ of coordinates $p_i$, we want to get the coordinates $u_i$ of an integer vector $\mathbf{u}$ that is such that $\mathbf{p}^t \mathbf{u} = 1$. Surprisingly, we could not find in literature algorithms in the cases $N > 3$. We propose here two recursive algorithms among the four we could find. They give different solutions that are all valuable.

Method-0. We consider $p_1$ and $p_2$ the two first coordinates of **p**, and we call $(u, v)$ their Bézout numbers, i.e. $up_1 + vp_2 = gcd(p_1, p_2)$. If we note $\{k_i, i = 2, ..., N\}$ the Bézout numbers in dimension $N - 1$ associated with the set $\{gcd(p_1, p_2), p_3, ..., p_N\}$, a solution of the initial $N$-dimensional Bézout's identity is $\{uk_2, vk_2, k_3, ..., k_N\}$. This method is easy to compute by recursion until the dimension decreases down to $N = 2$ for which the solution is given by the classical Bézout's algorithm. The problem related to this method is that the absolute values of the Bézout numbers $u_i$ can be quite high. One could screen all the pairs $(p_i, p_j)$ in place of $(p_1, p_2)$ to determine the lowest Bézout numbers but this method would be unrealistic for high dimensions $N$. We could find another method for which the values are lower than those usual found by method-0.

Method-1. The set of integers $\{p_i, i = 1,2, ..., N\}$ is sorted in the decreasing order of the absolute values. The sorting permutation σ is kept in memory. The smaller non-null value is called $p_{i_0}$. We calculate the quotient and residue sets $\{q_i, i < i_0\}$ and $\{r_i, i < i_0\}$ with $q_i = \left\lfloor \frac{p_i}{p_{i_0}} \right\rfloor$ and $r_i = p_i - q_i p_{i_0}$, the quotient and remainder of the Euclidean division by $p_{i_0}$. If we note $\{u_1, u_2, ..., u_{i_0-1}, u_{i_0}, 0, ..., 0\}$ the Bézout numbers associated with the set $\{r_1, r_2, ..., r_{i_0-1}, p_{i_0}, 0, ..., 0\}$, a solution of the $N$-dimensional Bézout's identity is $\{u_1, ..., u_{i_0-1}, u_{i_0} - \sum_{i=1}^{i_0-1} q_i u_i, 0, ..., 0\}$. This method is easy to compute by recursion until the dimension decreases down to $N = 2$ for which the solution is given by the classical Bézout algorithm. The correct order of the Bézout numbers associated with the initial set $\{p_i, i = 1,2, ..., N\}$ is restored by applying $\sigma^{-1}$. The Bézout numbers calculated with this method are smaller in absolute value than those obtained by method-0. For example, with the vector $\mathbf{p}^t$ = [51, 450, −102, 240, −277, 54, 450, 532], method-0 gives $\mathbf{u}^t$ = [−4876, 552, 0, 0, −1, 0, 0, 0], and method-1 $\mathbf{u}^t$ = [−3, 0, 0, 0, 0, −3, 0, 1]. The calculation lasts only a few ms. Even if method-1 gives small Bézout vectors **u**, it may not give systematically the smallest ones. We will see in the next section, how to calculate shorter Bézout vectors **u** with the help of the column-constrained unimodular matrices determined in §2.

## 4. Hyperplane unit cell

Let us assume that a hyperplane is given only by its normal vector **p**, an integer vector of the reciprocal space. In 3D crystallography, we would say that we know the Miller indices $h, k, l$ of the plane **p**, i.e. $\mathbf{p}^t = (h, k, l)$. How to determine a unit cell such that $N - 1$ vectors of this cell belong to the hyperplane, and one vector is in the first layer? In other words, we are looking for $N$ vectors $\{\mathbf{b}_1, ..., \mathbf{b}_j, ..., \mathbf{b}_N\}$ such that $N - 1$ vectors $\{\mathbf{b}_2, ..., \mathbf{b}_j, ..., \mathbf{b}_N\}$ are such that $\mathbf{p}^t \mathbf{b}_j = 0$ (layer $q = 0$), and the first vector $\mathbf{b}_1$ is such that $\mathbf{p}^t \mathbf{b}_1 = 1$ (layer $q = 1$). We propose a method to calculate a unit cell based on the solution of the column-constrained unimodular matrix problem described in §2.

We start from the input vector **p**. A Bézout vector $\mathbf{b}_1$ in the layer $q = 1$ is given by the Bézout vector associated with **p** by the algorithm detailed in §3 (method-1). Now, how to determine the $N - 1$ vectors in the layer $q = 0$? We consider the unimodular matrix **M** that is such that the last column is the vector $\mathbf{b}_1$. The $N - 1$ first column vectors of the matrix **M** are called $\mathbf{v}_j$ for $j \in \{2,..N\}$. Each of these vectors belongs to the lattice; thus, they are such that $\mathbf{p}^t \mathbf{v}_j = q_j \in \mathbb{Z}$. We construct by parallel project along $\mathbf{b}_1$ the new vectors $\mathbf{b}_j = \mathbf{v}_j - q_j \mathbf{b}_1$. They verify $\mathbf{p}^t \mathbf{b}_j = 0$ for $j \in \{2,..N\}$; i.e. they belong to the layer $q = 0$. A 3D example of these geometrical projections parallel to $\mathbf{b}_1$ is represented in Figure 1.

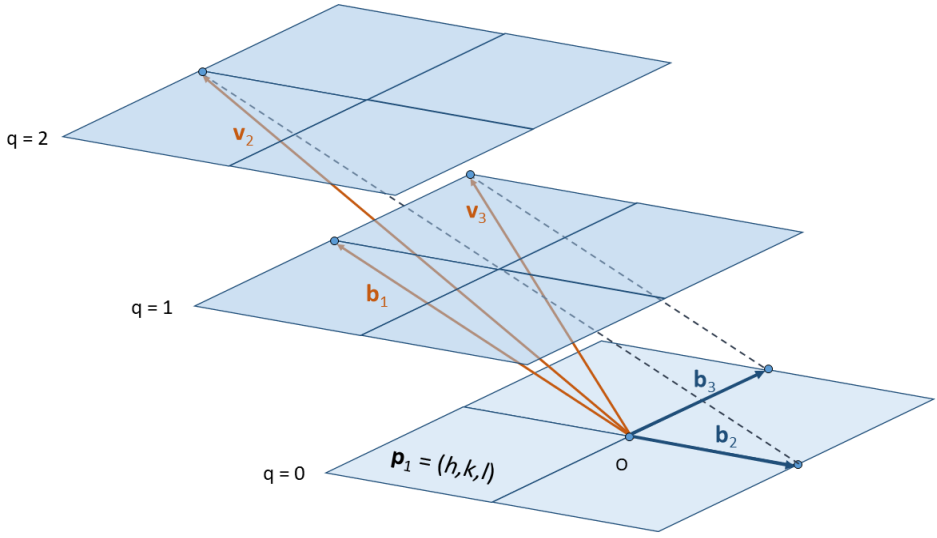

**Figure 1** Construction of the unit cell attached to the hyperplane $\mathbf{p}_1 = (h, k, l)$. First, the vector $\mathbf{b}_1$-such that $\mathbf{p}_1^t \mathbf{b}_1 = 1$ is found by the Bézout's algorithm detailed in §3. Then a unimodular matrix $\mathbf{M} = (\mathbf{v}_2, \mathbf{v}_3, \mathbf{b}_1)$ is found with the algorithm described in §2. The vectors $\mathbf{b}_2$ and $\mathbf{b}_3$ are obtained by projections along the vector $\mathbf{b}_1$. The unit cell attached to the hyperplane $\mathbf{p}_1$ is $(\mathbf{b}_2, \mathbf{b}_3, \mathbf{b}_1)$.

We get a unit cell $\mathbf{U} = (\mathbf{b}_1, ..., \mathbf{b}_j, ..., \mathbf{b}_N)$ attached to the plane **p** such that $det(\mathbf{U}) = 1$, $\mathbf{p}^t \mathbf{b}_1 = 1$, and $\mathbf{p}^t \mathbf{b}_j = 0$ for $i \in \{2,..N\}$. It is the unit cell we were looking for. However, the vectors $\mathbf{b}_j$ may be quite large and their angles far from orthogonality. There are two methods to find a reduced unit cell $\mathbf{U}' = (\mathbf{b}'_1, ..., \mathbf{b}'_j, ..., \mathbf{b}'_N)$, with $\mathbf{b}'_j$ with shorter lengths and angles between each other's closer to orthogonality, and that have the same properties with the vector **p**. One could apply the LLL algorithm, from the names of its authors Lenstra-Lenstra-Lovász (1982), that is well-known in computer science and cryptography, or the alternative algorithm called cubification (Cayron, 2021b). The latter choice is chosen. We wrote a computer program in Python 3.8 called *GeneralizedBezout* that incorporates the hyperplanar reduction modules previously developed in the Python program *Cubification*. This module gives a reduced unit cell in a few ms on our standard 6 years old laptop computer.

Let us give an example with $\mathbf{p}^t = (-54, 131, -48, 632, 23, 177, 333, 99, -581, 377)$. The coordinates were chosen completely "randomly" by the author. The Bézout vector associated with the plane $\mathbf{p}$ given by method-1 described in §3 is $\mathbf{b}_1^t = [1, 0, 0, 0, 11, 0, 0, -2, 0, 0]$. After determining a first unit cell by projections parallel to $\mathbf{b}_1$, and after reducing this unit cell, this vector becomes $(\mathbf{b}_1')^t = [0, 1, 1, 0, 1, 0, 0, 1, 1, 1]$. The reduced unit cell is given by the matrix (the vectors are written in columns):

$$\mathbf{U}' = \begin{bmatrix} 0 & -1 & 1 & 1 & 0 & 0 & 0 & 0 & 0 & 0 \\ 1 & 0 & 0 & -1 & -2 & 1 & -1 & 0 & 0 & -1 \\ 1 & -2 & -2 & -1 & -2 & 0 & 1 & 1 & 0 & 1 \\ 0 & 0 & 0 & 0 & 0 & 0 & 0 & 0 & 1 & 0 \\ 1 & 0 & 1 & -1 & 1 & 2 & 0 & 0 & -1 & 1 \\ 0 & -2 & 1 & 0 & 0 & -1 & 0 & 1 & -1 & -1 \\ 0 & 0 & 1 & 1 & -1 & 0 & 0 & -1 & -1 & 1 \\ 1 & 0 & -2 & 0 & 1 & 0 & -2 & 0 & -1 & 0 \\ 1 & -1 & 0 & -1 & 0 & 0 & 0 & -1 & 0 & 0 \\ 1 & -1 & -1 & -2 & 1 & 0 & 1 & -1 & 0 & 0 \end{bmatrix}$$

The calculation was made in 20 ms.

The matrix $\mathbf{U}'$ is interpreted crystallographically / geometrically as the unit cell attached to the hyperplane $\mathbf{p}$. From an algebraic point of view, $\mathbf{U}' = (\mathbf{b}_1', ..., \mathbf{b}_i', ..., \mathbf{b}_N')$ can equivalently be understood as the infinite set of solutions of the $N$-dimensional Bézout's identity, where $\mathbf{b}_1'$ is a reduced solution of the equation $\mathbf{p}^t \mathbf{b}_1' = 1$, and the other vectors are such that $\mathbf{p}^t \mathbf{b}_j' = 0, j \in \{2, ..N\}$. The set of solutions of Bézout's identity are thus $\mathbf{b}_1' + \{\mathbb{Z} \, \mathbf{b}_j'\}$ with $j \in \{2, ..N\}$, where $\{\mathbb{Z} \, .\}$ means all the linear combinations with integer coefficients. This $N - 1$ dimensional affine space represents all the solutions of Bézout's identity made on the coordinates of $\mathbf{p}$.

We note that the solutions $\mathbf{b}_j'$ giving $\mathbf{p}^t \mathbf{b}_j' = 0$ could be of interest for other arithmetic problems. In general, they are given by an algorithm called PSLQ (Ferguson, Bailey & Arno, 1999; see also Wikipedia, 2021c), where "PS refers to partial sums of squares, and LQ to a lower trapezoidal orthogonal decomposition". The PSLQ algorithm works for any vector $\mathbf{p} \in \mathbb{R}^N$ and has permitted to discover numerous previously unknown identities among real numbers. One of them is the formula for the value of $\pi$ discovered by Bailey, Borwein & Plouffe (1997). Our algorithm gives only solutions for vectors $\mathbf{p} \in \mathbb{Z}^N$, but it may be more efficient. Let us consider the PSLQ is implemented in *Mathematica* under the function *FindIntegerNullVector*. When applied to the vector $\mathbf{p}^t = (-54, 131, -48, 632, 23, 177, 333, 99, -581, 377)$ given in the previous example, this function gives only one solution that is $[1,0,-2,0,2,0,2,0,0,-2]$. We notice that this vector is larger than all the vectors $\mathbf{b}'_j$ in columns $j \in \{2,\ldots N\}$ of the matrix $\mathbf{U}'$. This could mean that our algorithm may have some advantages over PSLQ to find integer relations between integers. More fundamental and applicative research is however required to confirm or infirm this point.

## 5. Conclusion

In a first step, an algorithm is proposed for the column-constrained unimodular matrix problem. It permits to find an integer matrix $\mathbf{M}$ such that $det(\mathbf{M}) = 1$ and the last column vector of $\mathbf{M}$ is a given vector $\mathbf{t}$. In a second step, a recursive algorithm that finds some solutions to the $N$-dimensional Bézout's identity is presented. To any integer vector $\mathbf{p}$, it gives an integer vector $\mathbf{u}$ such that $\mathbf{p}^t \mathbf{u} = 1$. At this stage, the vectors $\mathbf{u}$ may be large, and the infinite set of solutions is yet not determined. The third step combines the two previous ones and uses parallel projections to determine a unit cell attached to any integer hyperplane $\mathbf{p}$. This unit cell can be shorten by lattice reduction (LLL or cubification). Equivalently, the vectors of the unit cell form the $N-1$ affine space of all the solutions of the $N$-dimensional Bézout's identity made on the integer coordinates of the vector $\mathbf{p}$.

**Acknowledgements**    Prof. Roland Logé is warmly acknowledged for the freedom given to our research that sometimes goes beyond metallurgy.

**Note:** the Python program *GeneralizedBezout* will be made available for the reviewers of the present manuscript. If and once the manuscript accepted for publication, it will be deposited on github or freely available on demand.